\documentclass[prl,twocolumn,showpacs,floatfix,amsfonts]{revtex4}
\usepackage{graphicx,graphics,color,epsfig}
\usepackage{bm}
\usepackage{amsmath}
\usepackage{amssymb}
\begin{document}
\title{Quantum Nyquist Temperature Fluctuations
}

\author{A. V. Balatsky}
\affiliation{Theoretical Division, Los Alamos National Laboratory,
Los Alamos, New Mexico 87545}
\author{Jian-Xin Zhu}
\affiliation{Theoretical Division, Los Alamos National Laboratory,
Los Alamos, New Mexico 87545}

\begin{abstract}
{We consider the temperature fluctuations  of a small object.
Classical fluctuations of the temperature have been considered for
a long time. Using the Nyquist approach, we show that the
temperature of an object fluctuates when in a thermal contact with
a reservoir. For large  temperatures or large specific heat of the
object $C_v$, we recover standard results of classical
thermodynamic fluctuations $\langle \Delta T^2\rangle = \frac{k_B
T^2}{C_v} $. Upon decreasing the size of the object, we argue,
one necessarily reaches the quantum regime that we call quantum
temperature fluctuations. At temperatures below $T^{*}\sim
\hbar/k_{B}\tau$, where $\tau$ is the thermal relaxation time of
the system, the fluctuations change the character and become
quantum. For a nano-scale metallic particle in a good thermal
contact with a reservoir, $T^{*}$ can be on a scale of a few
Kelvin. }
\end{abstract}

\maketitle

Classical thermodynamic fluctuations have been studied for more
than two centuries. The Gibbs canonical distribution function,
$P(\alpha)d\alpha\propto \exp[-E(\alpha)/k_{B}T] d\alpha$, is one
of the most fundamental concept in statistical physics. All
thermodynamic variables can be obtained from this distribution
function. According to the initial assertion of Gibbs, the
temperature of a canonical ensemble is constant and thus does not
fluctuate. Therefore, the temperature fluctuation cannot be
generically represented by the above distribution function.
Instead it is derived from energy fluctuations ($\delta E$).
Alternatively, the von Laue approach~\cite{LL80} to fluctuating
system thermodynamics via the minimal work can also lead to the
temperature fluctuation ($\delta T$). In recent years, there has
been increasing interest in the nano-scale problems such as the
glass transition~\cite{KTW89}, nucleation~\cite{Ring2001}, and
protein folding~\cite{Matt00}. Of device importance, the
mechanical resonators are being pushed to the nanometer
scale~\cite{CR98,CR01}. For these nano-scale systems, the
temperature fluctuation can be large. Recently, it has been
shown~\cite{DHS00} that the von Laue approach gives much more
reasonable results for the temperature fluctuation in a confined
geometry than the treatment with the Gibbs distribution function.
To the best of our knowledge, the existing study has been limited
to the classical regime. We know that any classical variable, say
coordinate, force, etc., has its corresponding standard quantum
limit where quantum fluctuations dominate. Similarly, we expect
temperature $T$  will have its quantum limit. Here we argue that
when the temperature is below $T^{*} \sim \hbar/\tau$, where
$\tau$ is the thermal relaxation time of the nano-scale particle,
a quantum temperature fluctuation regime emerges.

\begin{figure}
\epsfig{figure=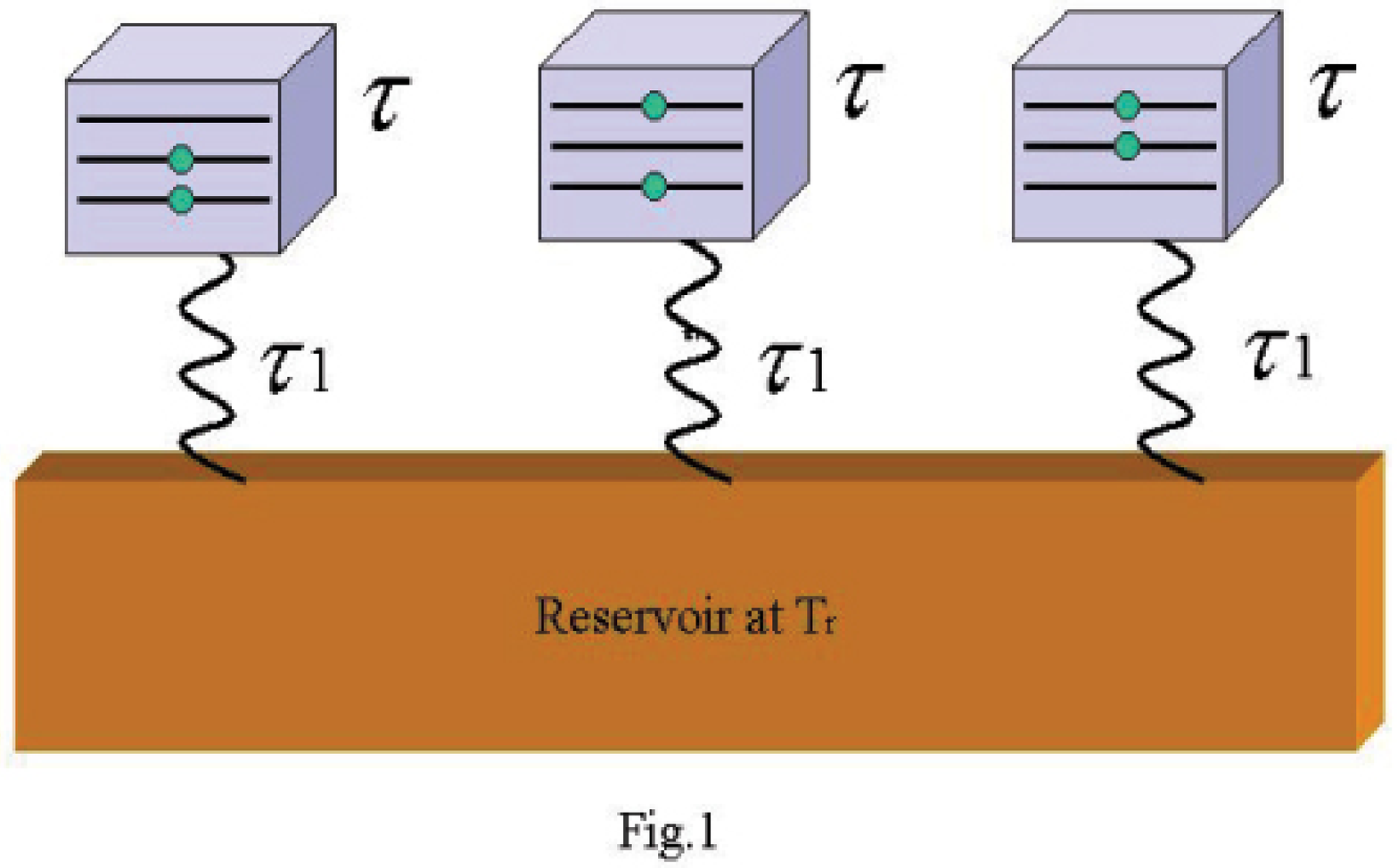,width=5 cm}
\caption[*]{ The ensemble of the quantum dots. The thermal
relaxation time within each dot is $\tau_1$. Each dot is thermally
coupled to the reservoir with temperature $T$. Note that each dot
contains large enough number of particles that the thermodynamics
consideration is still applicable. In particular the notion of a
quasi-equilibrium distribution is applicable to each dot. Consider
a thermal fluctuation that produces the temperature fluctuation in
the $i$th dot, $T+\delta T_{i}$. After the equilibration time
$\tau_1$, the dot relaxes to the quasi-equilibrium state with
temperature $T$. Repeating this measurement on a set of dots, we
find the distribution of temperature in an ensemble of dots. The
width of this distribution of $T$ will determine the fluctuation
of temperature $\delta T$. } \label{FIG:DOTS}
\end{figure}

Consider an experiment in which we are going to measure the
temperature fluctuation of an ensemble of increasingly small
objects. Without loss of generality consider a set of quantum
dots, as shown in Fig.~\ref{FIG:DOTS}. Assume these dots to be
similar in the number of contained particles, size, etc. In
addition, each dot has discrete levels, which are filled by a
sufficient number ($N\gg 1$) fermions (e.g., electrons) or bosons
(e.g., $^{4}$He atoms). All these dots are in contact with a
substrate (large plate) which plays the role of a thermal
reservoir. The reservoir is kept at a certain temperature $T$. The
thermal contact between the dots and the reservoir will cause the
thermal fluctuations in the dots. As a result, the heat flows
to/from the reservoir. The relaxation time for the thermal process
between the dots and the reservoir is $\tau$.

Let us first briefly recall the derivation of the temperature
fluctuation from the von Laue approach~\cite{LL80}. The
propability $w$ for a fluctuation is proportional to
$\exp(-R_{min}/k_{B}T)$, where $R_{min}$ is the minimum work
needed to fulfill reversibly the given change in the thermodynamic
quantities in the quantum dot and $k_{B}$ is Boltzmann's constant.
For simplicity, we assume the volume of each dot be fixed so that
$R_{min}=\Delta E-T\Delta S$, where $\Delta E$ and $\Delta S$ are
respectively the changes in the energy and entropy. Therefore we
have:
\begin{equation}
w=\mathcal{N}_{0} \exp[-\frac{\Delta E-T\Delta S}{k_{B}T}]\;,
\label{EQ:PROBABILITY}
\end{equation}
where $\mathcal{N}_{0}$ is the normalization constant. For small
fluctuations, by expanding $\Delta E$ to the second order in $\Delta S$,
and noticing that $\Delta S=(C_{v}/T)\Delta T$ with $C_{v}$ the
specific heat, it is found:
\begin{equation}
w=\mathcal{N}_{1} \exp[-\frac{C_{v}\Delta T^{2}}{2k_{B}T}]\;,
\end{equation}
where $\mathcal{N}_{1}$ is a new renormalization constant.
It follows immediately that the average square fluctuation of
temperature at a constant volume is:
\begin{equation}
\langle \Delta T^{2}\rangle=T^{2}/C_{v}\;.
\end{equation}
This is the result from the standard classical theory of
thermodynamic fluctuations. However, it has to be kept in mind
that for the starting equation~(\ref{EQ:PROBABILITY}) to be valid,
the temperature has to be much bigger than the thermal relaxation
rate, i.e.:
\begin{equation}
k_{B}T\gg \hbar/\tau\;,
\end{equation}
which also means when the temperature is too low or $\tau$ is too
small, the fluctuations can no longer be treated classically.

Here we propose the use of the Nyquist approach~\cite{Nyquist28}
to treat the temperature fluctuation.  First consider a
generalized coordinate $x$ and its relaxation $\frac{dx}{dt}=
-\lambda(x-x_0)$, where $x_0$ is its equilibrium value. The
external force $F$, conjugated to the coordinate $x$, determines
the equilibrium $x_0$. Now, in case there are fluctuations in the
external force $F$, equilibrium value $x_0 = {\bar x}_0 + \Delta
x_0$ fluctuates around its steady state position ${\bar x}_0$ by $
\Delta x_0 = \frac{\partial x}{\partial F}|_{{\bar x}_0} \Delta
F$. For the equation of motion we get
\begin{equation}
\frac{dx}{dt} = -\lambda(x-{\bar x}_0 - \frac{\partial x}{\partial
F}|_{{\bar x}_0} \Delta F) \label{EQ:genforce}
\end{equation}
Assume now   temperature $T = x$ to play the role of a generalized
coordinate and the entropy $S = F$ the role of the generalized
fluctuating force. The relaxation process of the temperature can
be described by a linearized macroscopic ``equation of motion'':
\begin{equation}
\frac{d\Delta T}{dt}=-\lambda (\Delta T -  \frac{\partial
T}{\partial S}|_{{\bar T}_0} \Delta S)\;,
 \label{EQ:RELAXATION}
\end{equation}
where $\lambda=1/\tau$,  and $\Delta T = T-\bar{T}_{0}$, and is
the deviation of the  equilibrium temperature  $T$ as a result of
the fluctuating force $\Delta S$. Equation~(\ref{EQ:RELAXATION})
is valid for the positive time and can be extended to the negative
time by changing sign of the derivative. Performing the Fourier
transform for $\Delta T$
\begin{equation}
\Delta T(t)=\frac{1}{2\pi}\int d\omega \Delta T_{\omega}
e^{-i\omega t}\;,
\end{equation}
and similarly for $\Delta S$, we can arrive at
\begin{equation}
\Delta T_{\omega}=\alpha(\omega) \Delta S_{\omega}\;,
\end{equation}
where the response function or generalized susceptibility reads:
\begin{equation}
\alpha(\omega)=\frac{\lambda T}{C_{v}(-i\omega+\lambda)}\;,
\end{equation}
where we have used $\frac{\partial T}{\partial S}|_{{\bar T}_0} =
\frac{T}{C_v}$. Using the fluctuation-dissipation theorem as
developed by Callen and  Welton, which relates the fluctuation of
a thermodynamic quantity to the imaginary part of the
susceptibility $\alpha (\omega)$~\cite{LL80,Callen51,GP87}, it
immediately follows:
\begin{equation}
\langle \Delta T^{2}\rangle_{\omega} =\hbar \coth(\hbar
\omega/2k_{B}T) \alpha^{\prime\prime}(\omega)\;,
\label{EQ:TOMEGA}
\end{equation}
where the imaginary part of $\alpha(\omega)$:
\begin{equation}
\alpha^{\prime\prime}(\omega)=\frac{\lambda
T}{C_{v}}\frac{\omega}{\omega^{2}+\lambda^{2}}\;.
\end{equation}
For the average quadratic fluctuation of $T$, it can be found:
\begin{equation}
\langle \Delta T^{2}\rangle=\frac{\hbar \lambda T}{2\pi C_{v}}
\int_{-\infty}^{\infty} d\omega
\frac{\omega}{\omega^{2}+\lambda^{2}} \coth(\hbar
\omega/2k_{B}T)\;. \label{EQ:FLUCTUATION}
\end{equation}
The integral on the right side of Eq.~(\ref{EQ:FLUCTUATION})
depends on the ratio of $k_{B}T /\hbar \lambda$. When $k_{B}T
/\hbar \lambda\gg 1$, by expanding $\coth(\hbar \omega/2k_{B}T)
=2k_{B}T/\hbar \omega +\hbar \omega/6k_{B}T$ for $\hbar
\omega/2k_{B}T\ll 1$, we have
\begin{equation}
\langle \Delta T^{2}\rangle =\frac{k_{B}T^{2}}{C_{v}} \left[
1+\frac{\hbar \lambda}{\pi k_{B}T}\ln\frac{\hbar
\omega_{c}}{k_{B}T}\right]\;, \label{EQ:TFLUC1}
\end{equation}
where we have introduced an upper band cutoff $\omega_{c}\sim
1/\tau_1$ on the order of the relevant bandwidth since at the high
frequency the integral $\int_{0}^{\omega_{c}} d\omega
\frac{\omega}{\omega^{2}+\eta^{2}}=\ln (\omega_{c}/\eta)$ is
logarithmically divergent as $\omega_{c} \rightarrow \infty$. One
can recognize immediately that Eq.~(\ref{EQ:TFLUC1}) is the
classical limit of the temperature fluctuations as derived above
from the von Laue approach. In the opposite limit of low
temperatures, $\hbar \lambda \gg k_{B}T$,  one finds:
\begin{equation}
\langle \Delta T^{2}\rangle =\frac{\hbar \lambda T}{\pi C_{v}} \ln
\frac{\omega_{c}}{\lambda}\;. \label{EQ:TFLUC2}
\end{equation}
Therefore, we find that at low temperatures the temperature
fluctuations would acquire a distinctly quantum character with
$\hbar /\tau$ entering into the magnitude of $\langle \Delta
T^{2}\rangle$. From ergodicity assumption, it follows that the time
averaged temperature of each particular dot is equal to the average value 
of $T$. Any fluctuation, described by Eq.(\ref{EQ:TFLUC2}), happen on a
characteristic time scale $\tau$.

The high temperature expansion in Eq.~(\ref{EQ:TFLUC1}) has
already indicated the crossover temperature
\begin{equation}
T^{*}=\frac{\hbar \lambda}{k_{B}\pi} \ln
\frac{\omega_{c}}{\lambda}\;,
\end{equation}
at which there is a change of the regime from the classical to
quantum fluctuations. Physically, $T^{*}\approx \hbar/k_{B}\tau$
corresponds to the uncertainty in energy associated with the
relaxation process in the subsystem. The reservoir is attached to
a subsystem via a thermal contact that has its own bandwidth
$\hbar /\tau$ and any temperature fluctuation will relax on the
scale of $\tau$. Once $T\ll T^{*}$, the intrinsic bandwidth of the
contact rather than the temperature will dominate the Gaussian
fluctuations. As stressed in Ref.~\cite{LL80}, fluctuations cannot
be treated classically if a fluctuating quantity is changing too
rapidly or if the subsystems are too small.

Our derivation, which is essentially identical to the classical to
quantum Nyquist crossover in the standard Nyquist theory, provides
the description of a quantum regime of the temperature
fluctuations.  There are few possible limitations of this
description. One is that our approach in not applicable to the
systems outside of thermal equilibrium, say hot electrons and cold
phonon bath, as sometimes is the case. Another restriction is that
the typical energy level spacing $\delta \sim 1/L^{d}$, $d$ is the
dimensionality of the dots, should be small compared to $T$. If the
temperature is much smaller than the level spacing, there will be
no thermally excited states and the notion of the thermodynamic
equilibrium is not applicable to this system. We should also point
out that for small particles there are mesoscopic corrections to
the total energy of particle, that can be related to the
temperature fluctuations. These mesoscopic fluctuations have a
typical temperature scale given by Thouless energy $k_B T_{Th} =
D/L^2$, $L$ - is the typical size of the particle, $D$- is
diffusion coefficient, and occur in addition to the fluctuations
we consider here. We assume here that $T^* > T_{Th}$.

\begin{figure}
\centerline{\psfig{figure=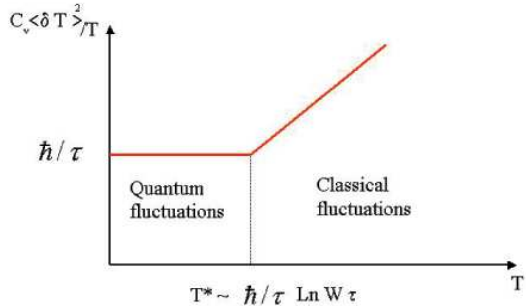,height=6cm,width=8cm,angle=0}}
\caption[*]{Temperature dependence of the fluctuations $\delta T$.
As $T\gg T^{*}$, the classical regime is recovered. As the
temperature gets smaller or the subsystem is getting smaller, a
quantum fluctuation regime comes into play. Note that the specific
heat $C_{v}$ ($\sim T$ for fermions and $\sim T^{3}$ for phonons
at low temperature) is temperature dependent. }
\label{FIG:FLUCTUATION}
\end{figure}

Experiments on the temperature-dependent fluctuations of
magnetization of small paramagnet were performed by Chui {\em et
al.}~\cite{Chui92}. Experimentally observed spectral density
$\langle \Delta T^{2}\rangle_{\omega}$ was shown to be of the form
given by the high temperature expansion of Eq.~(\ref{EQ:TOMEGA}).
The thermal relaxation time was $\tau \sim 1$ second for the
considered size of the paramagnet (about 1 cm$^{3}$). The total
temperature fluctuation $\langle \Delta T^{2}\rangle=T^{2}/C_{v}$
was also claimed as a result of the integration over
Eq.~(\ref{EQ:TOMEGA}). Therefore we can regard this result as an
experimental evidence for the classical temperature fluctuation in
the canonical ensemble according to Eq.~(\ref{EQ:TOMEGA}). The
obvious next step is to extend these measurements to the samples
of much smaller sizes, down to 1 $\mu$m in its linear size and
study the temperature dependence of the fluctuation spectrum at
low temperatures.

Now let us  estimate the crossover temperature $T^{*}$ for a
metallic dot and for a $^{4}$He droplet. The low temperature limit
$T\ll \hbar\lambda/k_{B}$ implies that the relaxation time of the
thermal object has to be short enough. Since the thermal
relaxation time $\tau=C_{v}R_{T}$, where $R_{T}$ is the thermal
resistance of the contact between the object and the thermal
reservoir. The easiest way to achieve a short thermal relaxation
time is to work with the smaller objects (smaller $C_{v}$) with
good enough thermal contact with the reservoir.

(i) For the metallic system, we consider a nanometer mechanical
resonator, which is a cylindrical gold (Au) rod of 1 $\mu\mbox{m}$
in length $L$ and 15 nm in radius $r$. The mass of this Au rod is
$m_{Au}=\rho_{Au} L\mathcal{A}=1.36\times 10^{-14}\;\mbox{g}$, where the
mass density of Au $\rho_{Au}=19.3 \mbox{g}/\mbox{cm}^{3}$, and
$\mathcal{A}=\pi r^{2}$ is the cross-section area. The mole mass
of Au is 196.97 g/mole, we thus find the specific heat constant
for Au per gram is $\gamma_{Au}=0.73 \mbox{mJ}/\mbox{mol}\cdot
\mbox{K}^{2}= 3.7\times 10^{-6} \mbox{J}/\mbox{g}\mbox{K}^{2}$. At
$T=0.1$ K, $C_{v}^{Au}=m_{Au}\gamma_{Au}=4.76\times 10^{-21}
\mbox{J}/\mbox{K}$. The thermal conductivity of Au at
$T=0.1\;\mbox{K}$ is about $\kappa_{T}=1$ Watt/K m. The thermal
resistance is $R_{T}=\frac{1}{\kappa_{T}}\frac{L}{\mathcal{A}}
=1.4\times 10^{9} \mbox{K}\cdot\mbox{Second}/\mbox{J}$. One then
finds $\tau\sim 6.7 \mbox{psec}$, and $T^{*}\sim 1$ K,
respectively, which is now experimentally accessible. In practice,
the thermal impedance mismatch at the interface between the
nano-scale subsystem  and the reservoir would lead to a much lower
conductance and a longer $\tau$. To estimate the role of the
``bottleneck'', one would need a specific model. However, from the
above estimate, one gets an impression that, regardless of the
experimental constraints, the quantum temperature fluctuation
below $T^{*}$ is observable.

(ii) For the case of a small bosonic system, we consider a droplet
of $^{4}$He, which is enclosed in a metallic container such as
lead. The size of the droplet is taken to be $0.1\;\mu\mbox{m}$.
Below the superfluid transition temperature, the specific heat of
$^{4}$He is dominated by phonons, which follows the power law as
$0.02\times T^{3} \mbox{J}/\mbox{g}\mbox{K}^{4}$~\cite{Wilks87}.
For the above given size of the droplet and at $T\sim 0.3$ K, the
specific heat $C_{v}=2.7\times 10^{-18} \mbox{J}/\mbox{K}$. At
these temperatures, the thermal resistance is dominated by a
surface resistance due to the contact of the droplet with the
metal. From Fig.~8.6 in Ref.~\cite{Wilks87}, we estimate the
thermal resistivity $R_{T} \sim 2
\mbox{K}\mbox{cm}^{2}/\mbox{Watt}$ between the $^{4}$He droplet
and the metal surface. One then obtains the relaxation times $\tau
\sim 6.9 \times 10^{-8}$ second, which corresponds to $T^{*}\sim
10^{-3}\mbox{K}$, which is small for the given size of the
droplet.

Experimentally the proposed crossover to quantum regime can be
seen as a change in temperature dependence of noise of some
observable. The choice depends on a specifics of the experiment
obviously, e.g for an oscillating  clamped beam \cite{CR01} it can
be a noise of the mechanical oscillatior. In the case of
magnetization noise \cite{Chui92}, one would desire to  measure
noise in the SQUID at relevant frequencies $1/\tau$.


In summary, we have used the Nyquist approach to study the
temperature fluctuation of an object in thermal contact with a
reservoir. It is shown for the first time that when at
temperatures below a characteristic value $T^{*}\sim
\hbar/k_{B}\tau$, the temperature fluctuation would acquire a
distinctly quantum character. For a nano-scale particle, $T^{*}$
is on the order of a few Kelvin. In light of recent advances in
nano-technology, the quantum fluctuation regime should be
experimentally accessible and might be relevant for the
experiments on nanoscale systems.

{\bf Acknowledgments}: We wish to thank B. Altshuler, J. Clarke,
J. C. Davis, S. Habib, H. Huang, A. J. Leggett,   R. Movshovich,
R. de Bruyn Ouboter, and B. Spivak for useful discussions.
This work was supported by the Department of Energy.

\end{document}